# Graphene films with large domain size by a two-step chemical vapor deposition process

*Xuesong Li[a], Carl W. Magnuson[a], Archana Venugopal[b], Jinho An[a], Ji Won Suk[a], Boyang Han[a], Mark Borysiak[c], Weiwei Cai[a], Aruna Velamakanni[a], Yanwu Zhu[a], Lianfeng Fu[d], Eric M. Vogel[b], Edgar Voelkl[d], Luigi Colombo[e*], and Rodney S. Ruoff[a*]*

[a]Department of Mechanical Engineering and the Texas Materials Institute, 1 University Station C2200, The University of Texas at Austin, Austin, TX 78712-0292

[b]Dept. Of Electrical Engineering, The University of Texas at Dallas

[c]2009 NNIN REU Intern at The University of Texas at Austin, Austin, TX 78712-0292

[d]FEI Company, 5350 NE Dawson Creek Drive Hillsboro, Oregon, 97124

[e]Texas Instruments Incorporated, Dallas, TX 75243

[*]Corresponding authors: r.ruoff@mail.utexas.edu (R.S.R.), colombo@ti.com (L.C.)

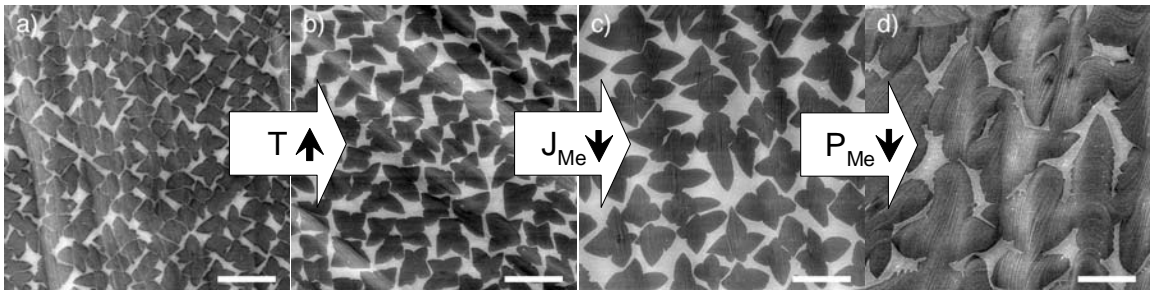

**Abstract**

The fundamental properties of graphene are making it an attractive material for a wide variety of applications. Various techniques have been developed to produce graphene and recently we discovered the synthesis of large area graphene by chemical vapor deposition (CVD) of methane on Cu foils. We also showed that graphene growth on Cu is a surface-mediated process and the films were polycrystalline with domains having an area of tens of square microns. In this paper we report on the effect of growth parameters such as temperature, and methane flow rate and partial pressure on the growth rate, domain size, and surface coverage of graphene as determined by Raman spectroscopy, and transmission and scanning electron microscopy. Based on the results, we developed a two-step CVD process to synthesize graphene films with domains having an area of hundreds of square microns. Scanning electron microscopy and Raman spectroscopy clearly show an increase in domain size by changing the growth parameters. Transmission electron microscopy further shows that the domains are crystallographically rotated with respect to each other with a range of angles from about 13 degrees to nearly 30 degrees. Electrical transport measurements performed on back-gated FETs show that overall films with larger domains tend to have higher carrier mobility, up to about 16,000 $cm^2\,V^{-1}\,s^{-1}$ at room temperature.

The fundamental properties of graphene are making it a very attractive material for use in electronics, optoelectronics, nano-electromechanical systems, chemical and bio-sensing, and many other applications.[1,2] Compared to other graphene synthesis processes reported to date, such as mechanical exfoliation of graphite,[3,4] reduction of graphene oxide,[5] or epitaxial growth on SiC substrates,[6,7] graphene growth on metal substrates[8-12] has the distinct advantage of being able to provide very large-area graphene films transferrable to other substrates. This advantage is especially true for the case of graphene growth on Cu substrates by chemical vapor deposition (CVD) of methane as reported by our group[12] and reproduced by other groups.[13,14] CVD of graphene on Cu yields a uniform graphene film whose size is limited only by the size of the Cu substrate and the growth system.

The presence of graphene on metal surfaces was observed as early as 1969 when the nomenclature "monolayer graphite" (MG) was first introduced by John May[15] in his rationalization of low energy electron diffraction (LEED) patterns that were, theretofore, unassigned. Later, Blakely and his research undertook extensive studies of mono- and bi-layer graphene on Ni substrates,[16-19] and several reviews (among others) of studies of growth of graphene on a wide variety of metal substrates are available.[20,21] It has been shown that graphene can grow across metal steps[10,22] and grain boundaries,[9,12] and graphene domains of a few hundred square microns have been observed on Ru substrates.[10] We have observed graphene domains of tens of square microns on Cu substrates using a C isotope labeling technique.[23] The C isotope labeling technique further demonstrated that graphene growth on Cu is surface-mediated, that is, C species decomposed from methane nucleate on the Cu surface and the nuclei grow to form islands and then domains (we note that some researchers may refer to domains as "grains") that cover the metal surface in its entirety.[12, 23] In our previous work we also observed that the domains might be in part defective as indicated by the appearance of D-band in the Raman spectra at certain locations. The inter-domain D-band defects are believed to arise from the misalignment of the domains as they come together to fully cover the Cu surface.[23] Because of the presence of such inter-domain defects we decided to try to develop processes to increase the domain size as an approach to decrease the density of such defects that are identifiable by the presence of the Raman D-band.

We studied the effect of growth parameters such as temperature (T), methane flow rate ($J_{Me}$), and methane partial pressure ($P_{Me}$) on the domain size of graphene grown on polycrystalline Cu. It should be noted that here we did not consider the effect of Cu grain orientations. Although graphene nuclei/islands on different Cu grains may have different shapes and densities, as reported previously,[12] at this time we believe that the shape of such islands may be controlled mainly by the growth conditions rather than the crystallographic orientations of the Cu substrate. We thus performed our investigation simply by comparing graphene islands with similar shape without concern about the specific Cu grain crystallographic orientation. In the case of the present parameter space, most of the graphene islands have a star-like shape and uniform density across the Cu foil, as shown by the scanning electron microscopy (SEM) images in Fig. 1 of partially grown graphene under different T, $J_{Me}$, and $P_{Me}$. In order to control the graphene domain density we performed experiments where we used the precursor flux and temperature to change the degree of supersaturation of active C-containing species (most probably $CH_x$) on the surface of the Cu to promote graphene nucleation. What we observed is that the density of graphene nuclei can be decreased (domain size increases) as T is increased (Figs. 1 a & b), or as $J_{Me}$ and $P_{Me}$ are decreased (Figs. 1 b & c) and (Figs. 1 c & d), respectively. That is, high T and low $J_{Me}$ and $P_{Me}$ were found to yield a low density of graphene nuclei and thus large domain size.

However, when $J_{Me}$ and $P_{Me}$ are less than a critical value, graphene does not nucleate and above these critical values graphene nuclei can form but, in a given range of values, graphene growth terminates before full surface coverage as shown in Fig. 1d. In this case the growth conditions under which partial coverage occurred were: growth temperature T = 1035 °C, methane flow rate $J_{Me}$ = 7 sccm and methane partial pressure $P_{Me}$ = 160 mTorr. Under these conditions even if the surface is continuously exposed to methane, full coverage is still not achieved. The precise growth conditions for partial coverage may vary from growth system to growth system. In order to fully cover the Cu surface the partial pressure of $CH_4$ must be increased; for example for $J_{Me}$ > 35 sccm and/or $P_{Me}$ > 500 mTorr in our growth chamber, complete surface coverage is obtained within 2 to 3 minutes.

The effects of $J_{Me}$ and $P_{Me}$ on graphene growth kinetics were further investigated by using the C isotope labeling technique together with *ex-situ* micro-Raman spectroscopy.[23] In these experiments the Cu surface was exposed to $^{13}CH_4$ and $^{12}CH_4$ (normal methane) sequentially. Since graphene growth on Cu occurs by surface adsorption, the isotope distribution in the local graphene regions will reflect the dosing sequence employed and can be mapped according to their different Raman mode frequencies.[23] For example, the G band of $^{13}C$ graphene is located at ~1520 cm$^{-1}$ while that of $^{12}C$ graphene is at ~1580 cm$^{-1}$ (Fig. 5g). By integrating the intensity of the G band of $^{13}C$ graphene over 1490-1550

cm$^{-1}$, the $^{13}$C graphene regions show as bright regions in the G$^{13}$-band maps while the $^{12}$C graphene regions show up as dark regions (Figs. 2 b and f); and vice versa for the G$^{12}$-band maps integrating the intensity of the G band of $^{12}$C graphene between 1550-1610 cm$^{-1}$ (Figs. 2 c and g). The regions that show up as dark in both maps are gaps uncovered by graphene, which can also be easily distinguished in the optical micrograph (e.g., Fig.2 a). As shown in Fig. 2, although the Cu surface was exposed to the alternating C isotopes a total of eight times, for the case of P$_{Me}$ = 160 mTorr, graphene growth terminated after the 6$^{th}$ dose with a maximum coverage of ~90%; whereas when P$_{Me}$ was increased to 285 mTorr, graphene growth terminated after the 4$^{th}$ dose and the surface was fully covered. The graphene growth rate was calculated by measuring the area of the isotopically labeled regions. Here we define the coverage rate, $v_{coverage}$, as the increase of graphene coverage (graphene area, $A_{graphene}$, divided by the total Cu area, $A_{Cu}$) within unit time ($t$):

$$v_{coverage} = \frac{dA_{graphene}}{A_{Cu}dt}$$

The average area growth rate of graphene domains, $v_{domain}$, is related to $v_{coverage}$ as

$$v_{domain} = v_{coverage}/n \qquad (2)$$

where $n$ is the domain (i.e., nucleus) density.

Fig. 3 shows a summary of the growth process of graphene on Cu at two different pressures, 285 mTorr and 160 mTorr. Figure 3a shows that as the Cu surface is exposed to methane at a P = 285 mTorr, the surface is fully covered with graphene after about 1.5 minutes whereas at a gas pressure of 160 mTorr the Cu surface reaches only 90% coverage after about 3 minutes and never reaches full coverage even after continued exposure to methane. Fig. 3b further shows that as the surface coverage increases or as the exposed Cu surface area decreases the graphene growth rate decreases dramatically.

We attribute the decrease in growth rate to the fact that the C species are supplied by the available Cu-catalyzed decomposition of methane.

In summary, graphene growth on Cu likely proceeds as follows:

1. Exposure of Cu to methane and hydrogen.
2. Catalytic decomposition of methane on Cu to form $C_xH_y$.
3. Depending upon the temperature, methane pressure, methane flow and hydrogen partial pressure, the Cu surface is either undersaturated, saturated, or supersaturated with $C_xH_y$ species.
4. Formation of nuclei as a result of local supersaturation of $C_xH_y$.
5. Nuclei grow to form graphene islands (in the case of an undersaturated surface, graphene nuclei do not form).
6. Full Cu surface coverage by graphene under certain T, P and J.

Therefore, there are three Cu surface conditions that can be considered: 1) undersaturated, 2) saturated and 3) supersaturated. In the case of an undersaturated Cu surface, graphene does not nucleate and no graphene is observed even though $C_xH_y$ may be present in the vapor phase and on the Cu surface. In the case of the saturated surface graphene, nuclei form, graphene grows to a certain island size and then it stops because the amount of $C_xH_y$ available from the exposed Cu surface is insufficient to continue driving the C attachment to the island edges and the Cu surface is only partially covered with graphene islands. That is, graphene islands, the Cu surface, and the vapor phase are in equilibrium. In the case where the Cu surface is supersaturated on the other hand, there is always enough methane to form sufficient $C_xH_y$ to drive the reaction between the $C_xH_y$ at the surface and the edges of graphene islands. In this last case graphene islands grow until neighboring islands connect to each other to fully cover the Cu surface.

Based on the above observations and the presence of defects at the inter-domain regions, as determined by the presence of a higher concentration of Raman D-band, we developed a two-step synthesis technique with the process flow shown in Fig. 4. For a given temperature (usually high temperature) nuclei are formed in step I at low $J_{Me}$ and $P_{Me}$ followed by step II, where the partial pressure of the methane is increased to promote full surface coverage. The resulting domains of the continuous graphene films grown by this technique were also delineated by C isotope labeling where $^{13}CH_4$ was used in step I followed by $^{12}CH_4$ (normal methane) in step II, as shown in Fig. 5. By comparing the $G^{12}$-band Raman maps, Figs. 5 b and e, it is clear that at low $J_{Me}$ and $P_{Me}$ the number of graphene nuclei is much smaller than that for high $J_{Me}$ and $P_{Me}$. As a consequence of the lower density of graphene nuclei, graphene can grow into larger domains. It is important to note that for a given temperature and partial pressure condition, once the nuclei density is set no significant new graphene nuclei are formed and subsequent changes in growth conditions only affect the graphene growth rate (we do not exclude that significant change in growth conditions may promote new nucleation). The Raman maps of D-bands (Figs. 5 c and f), a measure of defects in graphene, which are located at ~1310 cm$^{-1}$ and ~1350 cm$^{-1}$ for $^{13}C$ graphene and $^{12}C$ graphene, respectively, show that graphene films with larger domains have a lower density of defects arising from nucleation centers and inter-domain regions.

To further understand the domain structure and the impact of domains on electrical properties, we performed transmission electron microscopy (TEM) and electrical transport measurements using field effect transistor (FET). Figs. 6 a & b shows a high resolution TEM image and a mask filtered image of a monolayer graphene film at a domain boundary. As can be seen from the Fast Fourier Transform (FFT) images in Figs. 6c-e, the FFTs from the white and black box (Figs. 6c & d, respectively) in Fig. 6a show a single set of a hexagonal spot pattern rotated relative to one another in each image. The whole FFT of Fig. 6a shown in Fig. 6e shows the combination of the two FFT patterns rotated relative to one another by approximately 18.5°. The FFTs in Fig. 6c-e show clear evidence of the presence of domain boundaries in this image. Magnified images in Figure 6f-g of the graphene in the different domains show how the hexagonal graphene structure is oriented within the imaged graphene domain. Measurements on tens of domains show that the domain mis-orientation ranges from 13 to about 30 degrees (since graphene has a hexagonal structure, anything over 30 degrees is simply 60-X). Based on the TEM results together with the fact that the domain size of the films studied is much smaller than the Cu grain size (hundreds of microns to millimeters) and that graphene has a hexagonal structure and Cu is cubic it is likely that the graphene is not growing epitaxially on Cu except perhaps on (111) grains; we plan to further address this issue in the future.

We measured the transport properties of CVD graphene having two different average domain sizes, (linear length) of 6 µm and 20 µm, and compared them with those of graphene exfoliated from natural graphite. The effective mobility (µ) was extracted using the mobility model introduced by Kim *et al.*,[24] where the $R$–$V_{bg}$ curves were fit to give a constant value of mobility and intrinsic carrier concentration for each of the measured devices. Fig. 7 shows a summary of the transport properties of CVD graphene in comparison to exfoliated graphene. Fig. 7a shows the device layout of a back-gated FET where the light regions are Ni contacts and the dark region is the graphene. Fig. 7c shows a summary of the mobility of many devices for the three different types of graphene: graphene with 6 μm and 20 μm domains, and graphene exfoliated from natural graphite. The mobility for many devices from films with small domains, 6 μm, is in the range of ranges from 800 to about 7000 $cm^2V^{-1}s^{-1}$. In contrast, devices with larger graphene domains, 20 μm, have a much higher range of mobility, 800 to 16000 $cm^2V^{-1}s^{-1}$ compared to 2500 to over 40,000 $cm^2V^{-1}s^{-1}$ for exfoliated graphene. Obviously, even films with large domains have devices with low mobility. The lowest mobility for both 6 and 20 μm domain films may be associated not only with the presence of domain boundaries but also with the presence of wrinkles and other defects some of which are induced by the transfer process. The observation of high mobility in CVD graphene that is in the same range as that observed in exfoliated graphene suggests that we are improving the material quality based on our fundamental understanding of the graphene microstructure as well as its development during growth. There is still much work to be done before we have a complete understanding of the effect of structural defects and transport properties but we are beginning to develop the basic process variables that will help create higher quality graphene.

In summary, we investigated the effects of growth parameters such as temperature, methane flow rate and partial pressure on CVD synthesis of graphene on Cu substrates. High temperature and low methane flow rate and partial pressure are preferred to generate a low density of graphene nuclei, while high methane flow rate or partial pressure are preferred for continuous large-area graphene films. Based on these observations, we developed an isothermal two-step growth process in which a low graphene nuclei density is set followed by achieving full graphene surface coverage by increasing the methane flow rate and partial pressure. Electrical transport data showed that graphene films with large domains have higher mobility than those with small domains predominantly due to a decrease in inter-domain defects. Further TEM results indicate that the domains are rotated with respect to each other by as much as 30 degrees. The basic understanding presented in this paper can lead to significant improvements in graphene synthesis and graphene based electronic devices.

**Acknowledgements:**

The Ruoff group appreciates support from the Office of Naval Research, the Defense Advanced Research Projects Agency Carbon Electronics for RF Applications Center, and Nanoelectronic Research Initiative – SouthWest Academy of Nanoelectronics (NRI-SWAN) together with the Vogel group and L.C.

**Experimental**

*Graphene synthesis.* A split tube furnace with a 6-inch heating zone and a 1-inch outer diameter quartz tube with gas panel having methane and hydrogen was used in a CVD mode to grow graphene films on Cu. The graphene synthesis process presented here is similar to that previously reported[12] with small changes: the 25-μm thick Cu foil was first reduced and annealed at 1035 $^{o}$C under 2 sccm of $H_2$ at a pressure of 40 mTorr for 20 min to increase the Cu grain size and clean the Cu surface; the graphene growth temperature was changed to the desired value after the initial Cu cleaning and grain growth and a desired amount of methane was introduced into the growth tube; the methane partial pressure was controlled by either methane flow rate or a ball valve between the quartz tube outlet and the pump; the growth time was varied accordingly; the methane and hydrogen gas flow and pressure are kept constant, same values as the growth process, during the furnace cool-down.

*Micro-Raman characterization.* Graphene films were transferred onto 285-nm SiO$_2$/Si substrates using polymethyl-methacrylate (PMMA) [25] for optical microscopy and micro-Raman imaging spectroscopy (WiTec Alpha 300, 532 nm laser wavelength, 100x objective).

*Mobility Measurements.* Graphene films were transferred onto p-type Si wafers with a 300-nm thick SiO$_2$ layer using the same transfer method as previously[25] and FET devices were fabricated using standard photolithography and electron beam lithography processes; Ni was used for metal contacts. The channel lengths and widths ranged from 2 to 100 µm and 1 to 15 µm, respectively. The device structures were designed to permit multiple electrical measurements. All of the electrical measurements were performed using back gated devices at room temperature in air using a HP 4155 Semiconductor Parameter Analyzer.

*Transmission Electron Microscopy.* A Cs corrected FEI Titan S/TEM operating at 80keV was used for TEM imaging. HRTEM image was averaged with HoloWorks 5.0 using 20 individual images using an acquisition time of 0.5s per each frame for an improved signal/noise ratio. HRTEM was mask filtered using an array mask in Digital Micrograph. Magnifier function in Holoworks 5.0 was used to magnify cropped TEM images.

**References**

1. Geim, A. K.; Novoselov, K. S. *Nat. Mater.* **2007,** 6, 183-191.
2. Geim, A. K. *Science* **2009,** 324, (19), 1530-1534.
3. Novoselov, K. S.; Geim, A. K.; Morozov, S. V.; Jiang, D.; Zhang, Y.; Dubonos, S. V.; Grigorieva, I. V.; Firsov, A. A. *Science* **2004,** 306, 666-669.
4. Hernandez, Y.; Nicolosi, V.; Lotya, M.; Blighe, F. M.; Sun, Z.; De, S.; McGovern, I. T.; Holland, B.; Byrne, M.; Gun'Ko, Y. K.; Boland, J. J.; Niraj, P.; Duesberg, G.; Krishnamurthy, S.; Goodhue, R.; Hutchison, J.; Scardaci, V.; Ferrari, A. C.; Coleman, J. N. *Nat. Nanotech.* **2008,** 3, 563-568.
5. Park, S.; Ruoff, R. S. *Nat. Nanotech.* **2009,** 4, 217-224.
6. Berger, C.; Song, Z.; Li, X.; Wu, X.; Brown, N.; Naud, C.; Mayou, D.; Li, T.; Hass, J.; Marchenkov, a. A. N.; Conrad, E. H.; First, P. N.; de Heer, W. A. *Science* **2006,** 312, 1991-1996.
7. Emtsev, K. V.; Bostwick, A.; Horn, K.; Jobst, J.; Kellogg, G. L.; Ley, L.; McChesney, J. L.; Ohta, T.; Reshanov, S. A.; Rohrl, J.; Rotenberg, E.; Schmid, A. K.; Waldmann, D.; Weber, H. B.; Seyller, T. *Nat. Mater.* **2009,** 8, (3), 203-207.
8. Yu, Q.; Lian, J.; Siriponglert, S.; Li, H.; Chen, Y. P.; Pei, S.-S. *Appl. Phys. Lett.* **2008,** 93, 113103.
9. Reina, A.; Jia, X.; Ho, J.; Nezich, D.; Son, H.; Bulovic, V.; Dresselhaus, M. S.; Kong, J. *Nano Lett.* **2009,** 9, 30-35.
10. Sutter, P. W.; Flege, J.-I.; Sutter, E. A. *Nat. Mater.* **2008,** 7, 406-411.
11. Kim, K. S.; Zhao, Y.; Jang, H.; Lee, S. Y.; Kim, J. M.; Kim, K. S.; Ahn, J.-H.; Kim, P.; Choi, J.-Y.; Hong, B. H. *Nature* **2009,** 457, 706-710.
12. Li, X. S.; Cai, W. W.; An, J. H.; Kim, S.; Nah, J.; Yang, D. X.; Piner, R. D.; Velamakanni, A.; Jung, I.; Tutuc, E.; Banerjee, S. K.; Colombo, L.; Ruoff, R. S. *Science* **2009,** 324, 1312-1314.
13. Levendorf, M. P.; Ruiz-Vargas, C. S.; Garg, S.; Park, J. *Nano Lett.* **2009,** 9, (12), 4479-4483.
14. Lee, Y.; Bae, S.; Jang, H.; Jang, S.; Zhu, S. E.; Sim, S. H.; Song, Y. I.; Hong, B. H.; Ahn, J. H. *Nano Lett.* **2010,** 10, (2), 490-493.
15. May, J. W. *Surface Science* **1969,** 17, 267-279.
16. Shelton, J. C.; Patil, H. R.; Blakely, J. M. *Surf. Sci.* **1974,** 43, 493-520.
17. Eizenberg, M.; Blakely, J. M. *Surface Science* **1979,** 82, (1), 228-236.
18. Hamilton, J. C.; Blakely, J. M. *Surface Science* **1980,** 91, (1), 199-217.
19. Eizenberg, M.; Blakely, J. M. *Journal of Chemical Physics* **1979,** 71, (8), 3467-3477.
20. Oshima, C.; Nagashima, A. *J Phys-Condens Mat* **1997,** 9, (1), 1-20.
21. Wintterlin, J.; Bocquet, M.-L. *Surf. Sci.* **2009,** 603, 1841-1852.
22. Coraux, J.; N'Diaye, A. T.; Busse, C.; Michely, T. *Nano Letters* **2008,** 8, (2), 565-570.
23. Li, X. S.; Cai, W. W.; Colombo, L.; Ruoff, R. S. *Nano Lett.* **2009,** 9, 4268-4272.
24. Kim, S.; Nah, J.; Jo, I.; Shahrjerdi, D.; Colombo, L.; Yao, Z.; Tutuc, E.; Banerjee, S. K. *Appl. Phys. Lett.* **2009,** 94, 062107.
25. Li, X. S.; Zhu, Y. W.; Cai, W. W.; Borysiak, M.; Han, B.; Chen, D.; Piner, R. D.; Colombo, L.; S., R. R. *Nano Lett.* **2009,** 9, 4359-4363.

Table 1. Two-step graphene growth parameters

| Experiments | | T (°C) | $J_{Me}$ (sccm) | $P_{Me}$ (Torr) | t (min) | Average domain area ($\mu m^2$) |
|---|---|---|---|---|---|---|
| A | Step I | 1035 | 7 | 0.160 | 2.5 | 142 |
|   | Step II |   |   | 2 | 1 |   |
| B | Step I | 1035 | 35 | 0.460 | 0.5 | 33 |
|   | Step II |   |   | 2 | 1 |   |

Figure Captions.

**Figure 1.** SEM images of partially grown graphene under different growth conditions: T($^o$C)/J$_{Md}$(sccm)/P$_{Me}$(mTorr): (a) 985/35/460, (b) 1035/35/460, (c) 1035/7/460, (d) 1035/7/130. Scale bars are 10 μm.

**Figure 2.** Micro-Raman characterization of the isotope-labeled graphene grown under isothermal conditions, T (1035 $^o$C), and J$_{Me}$ (7 sccm) but different P$_{Me}$ (160 mTorr: panels a-d; 285 mTorr: panels e-h); (a) and (e) optical micrographs; (b) and (f) Raman maps G$^{13}$-band maps and (c and g) G$^{12}$-band maps corresponding to the area shown in (a and e), respectively; (d) and (h) schematically show the isotope distributions of the two cases in which the colors are decoded in the color bar with methane dosing sequences and times. Scale bars are 5 μm.

**Figure 3.** Graphical summary of graphene growth on Cu shown in Fig. 2. (a) Graphene coverage as a function of methane exposure time; (b) average graphene domain area growth rate as a function of coverage, respectively.

**Figure 4.** Two-step process flow of continuous graphene films with large domains.

**Figure 5.** Optical images and micro-Raman imaging spectroscopy maps of graphene films transferred onto 285-nm SiO$_2$/Si substrates. Optical images (a & d), Raman G$^{12}$-band maps (b & e) and D$^{12}$-band maps (c & f) of sample A (a-c) and B (d-f), respectively. (g) Raman spectra of $^{13}$C graphene, $^{12}$C graphene, $^{13}$C/$^{12}$C junction, wrinkle, nucleation center, and domain boundary, respectively, as marked with corresponding colored circles in (a-c). A summary of the two-step process for conditions A & B is shown in Table 1. Scale bars are 5 μm.

**Figure 6.** (a) TEM bright field image at 80keV of monolayer graphene (b) Mask filtered image of (a). (c)-(d) FFT from area in white and black box respectively in (a). FFTs in (c) and (d) show that the monolayer graphene has two different crystal orientations at each respective areas, and that the sample is not a non AB stacked bi-layer graphene. (e) FFT of the whole image in (a) shows two sets of hexagonal FFT spots mis-oriented by approximately 18.5 degrees from one another. (f)-(g) High resolution image cropped from white and black circled regions in (b) respectively.

**Figure 7.** (a) Optical micrograph of a FET device (top) and (b) a typical plot of the normalized channel resistance (R$_{ch}$) as a function of applied back gate voltage (V$_{bg}$). (c) Carrier mobility as a function of graphene domain size in comparison to exfoliated graphene.

Figure 1

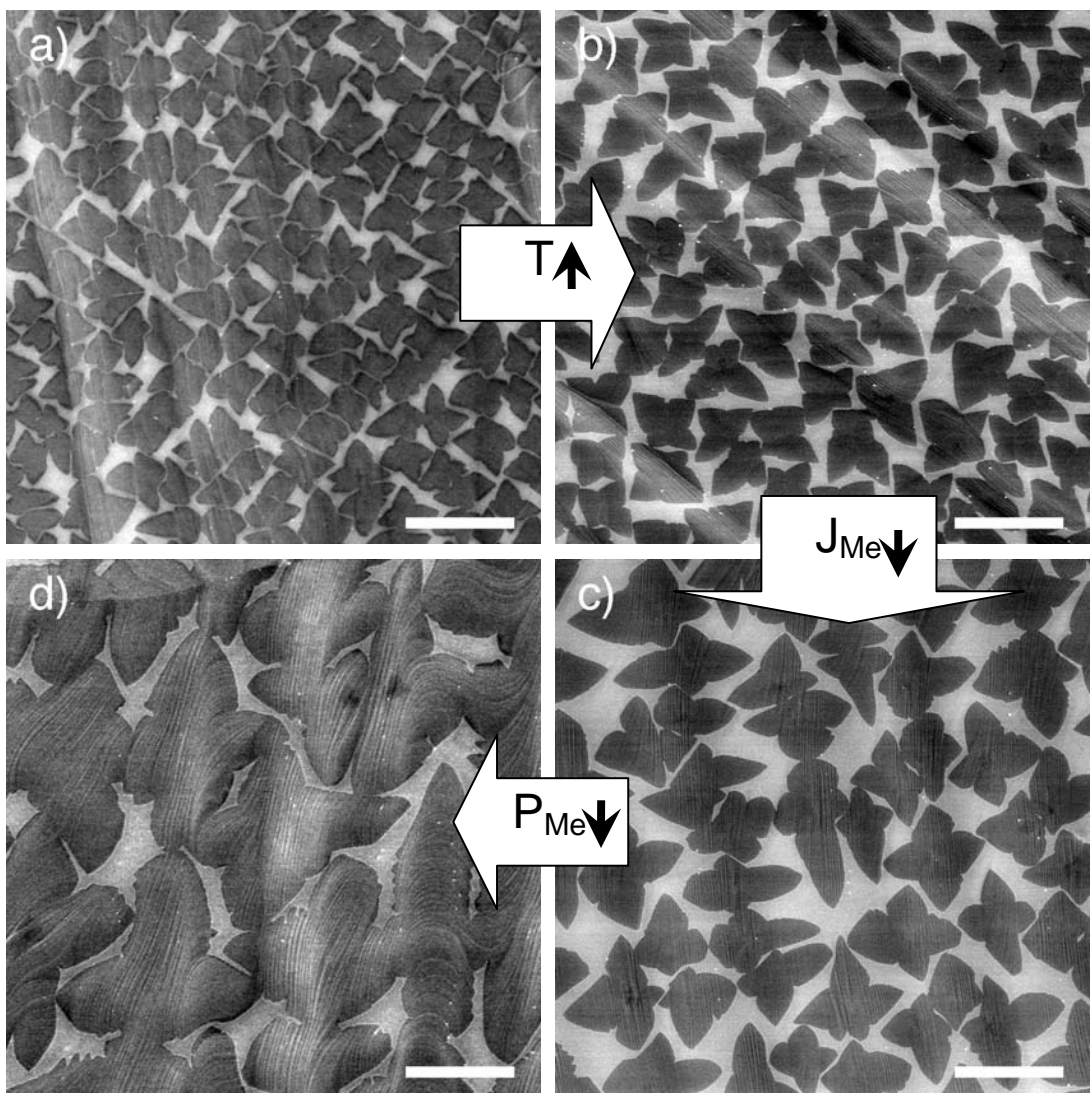

Figure 2.

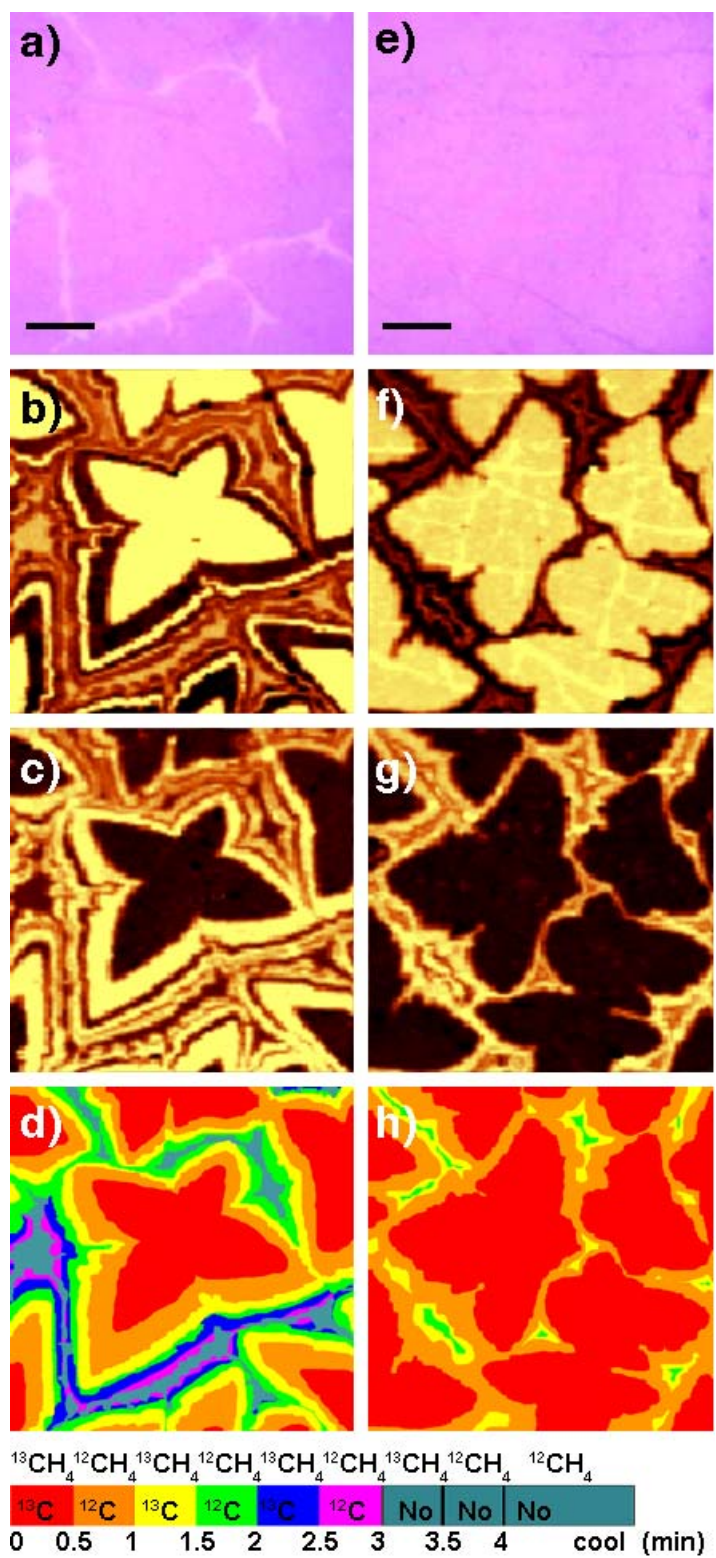

Figure 3

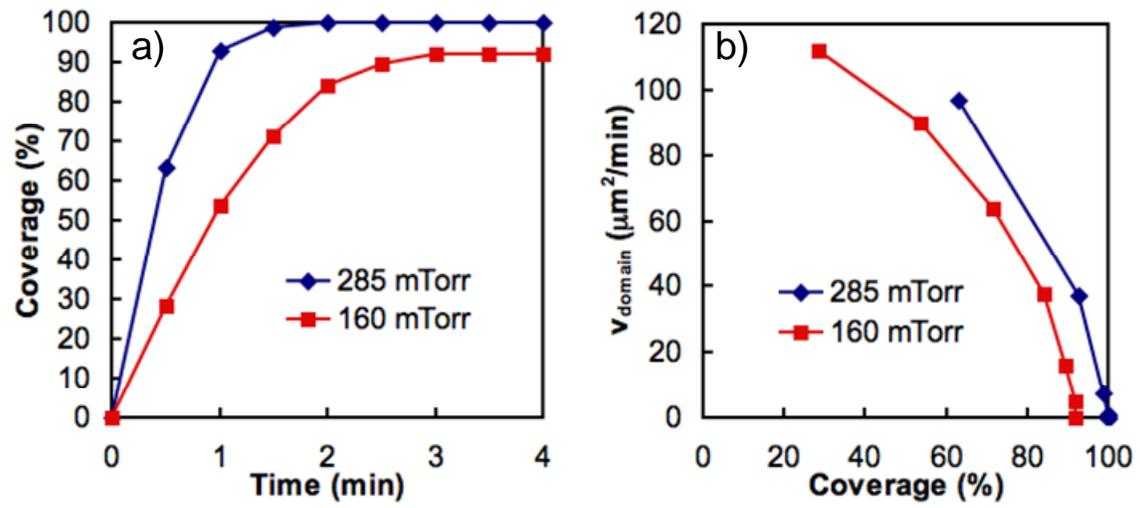

Figure 4

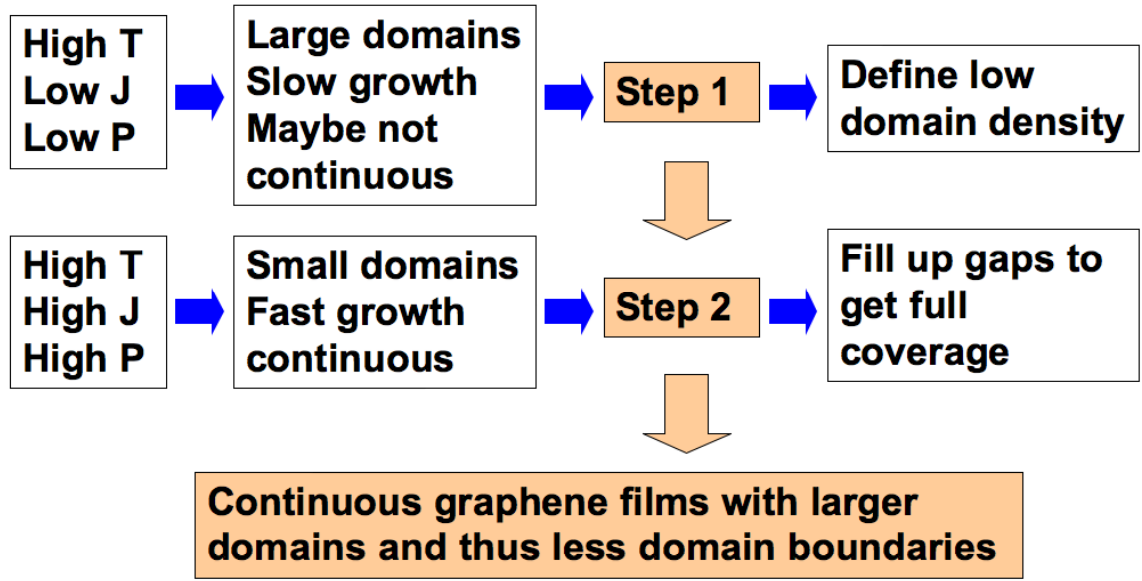

Figure 5

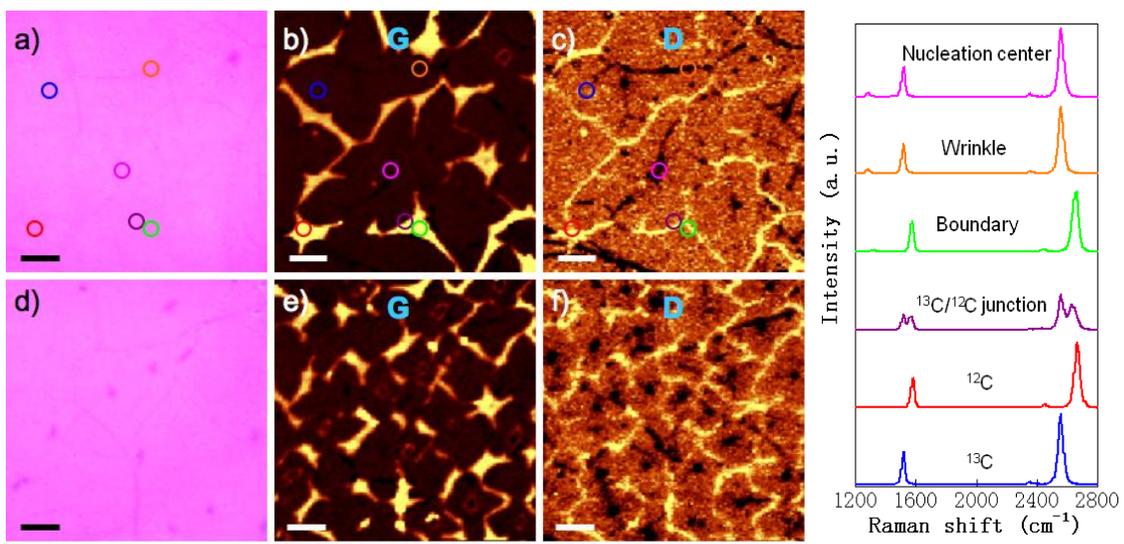

Figure 6

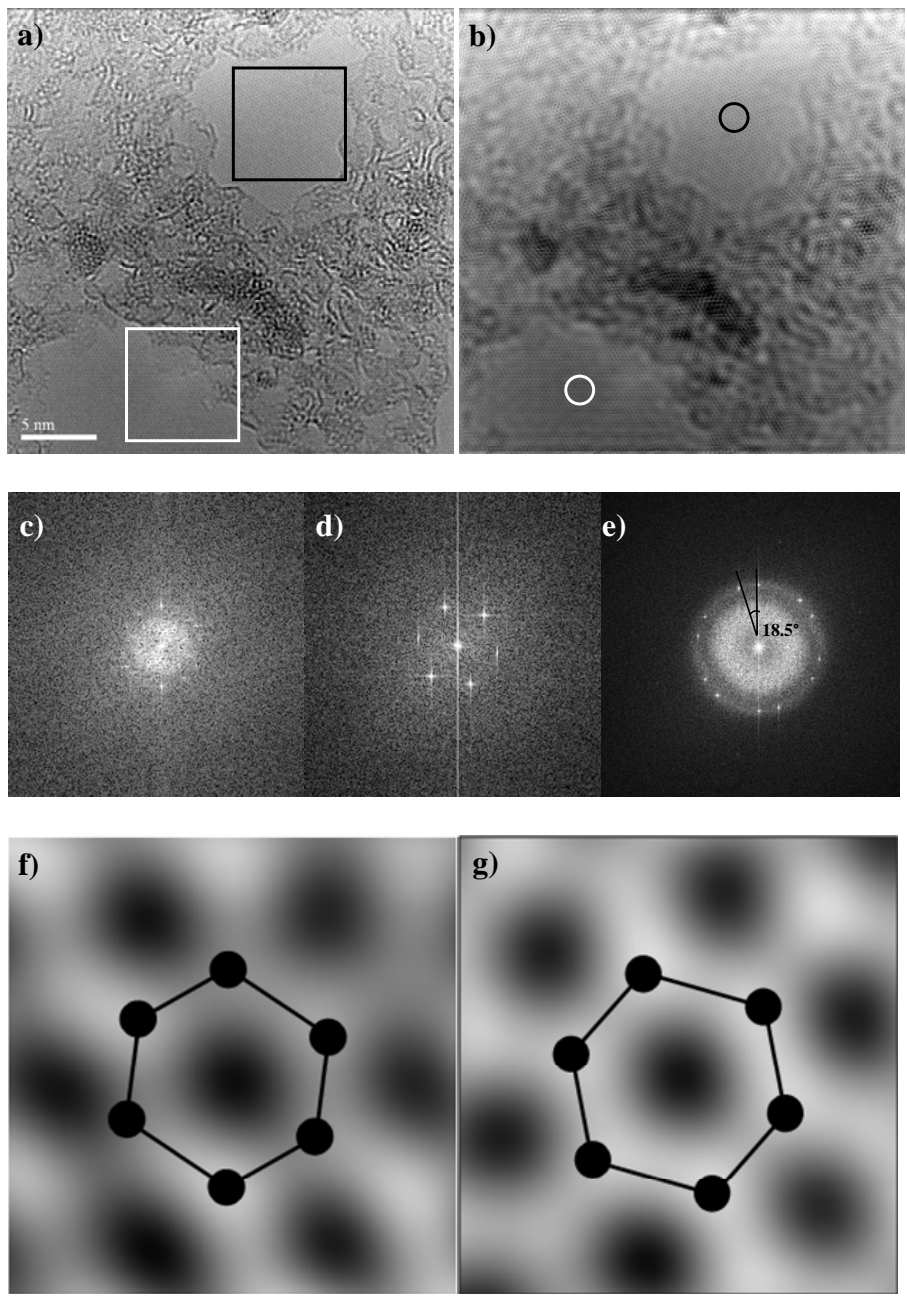

Figure 7.

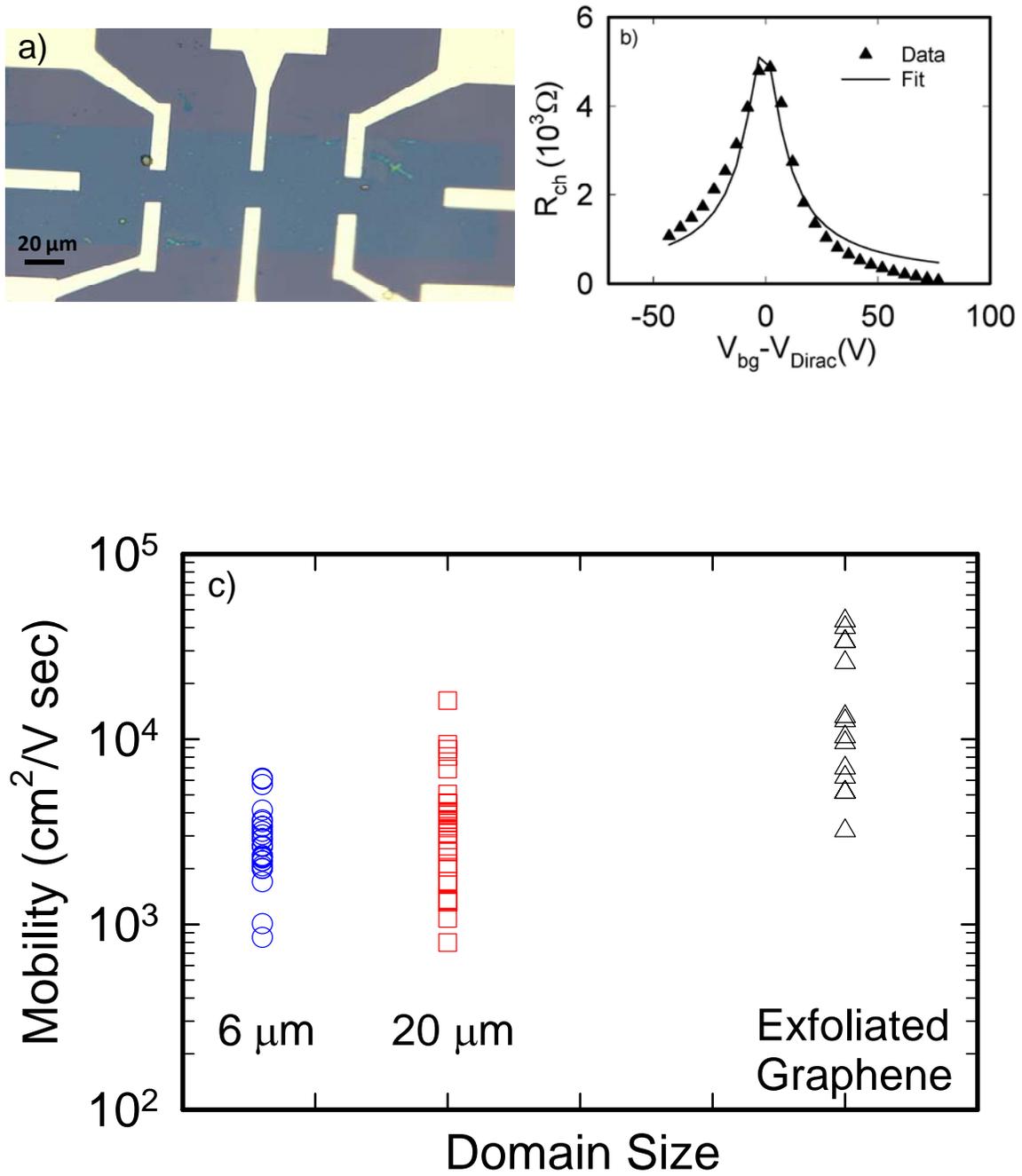